\documentclass[runningheads]{svmult}

\usepackage{makeidx}   
\usepackage{graphicx}  
\usepackage{subeqnar}  
\usepackage{multicol}  
\usepackage{physmubb}  
\makeindex             
\usepackage{epsfig}

\advance\voffset by 1.5in

\begin{document}
\title*{Why We Need Both the LHC and \\ an
$e^+e^-$ Linear Collider
\thanks{To appear in the {\it Proceedings of the 
14th Topical Conference on Hadron Collider Physics (HCP 2002)},
29 September --4 October, 2002, Karlsruhe, Germany.}
}

\author{S. Dawson\\
Physics Department, Brookhaven National Laboratory\\
Upton, N.Y., U.S.A.~~11973}

\authorrunning{S. Dawson}
\maketitle              

\section{Introduction}
The next high energy collider  will be the Large Hadron
Collider (LHC), to be completed in 2007.
The LHC, combined with RUNII at the Fermilab Tevatron, will greatly expand our
knowledge of physics at the TeV scale.
  In this note, we examine some of the physics
questions likely to be left unanswered by the LHC and discuss the role of
a high energy $e^+e^-$ linear
collider (LC) in providing answers.
We consider the connection between the understanding to
be found at the LHC and an LC, using Higgs physics, low energy supersymmetric
models, and the top quark sector as examples.

One of the central questions of particle physics is the origin of mass.
In the Standard Model, both fermion and gauge boson masses are generated
through interactions with a single scalar particle, the Higgs boson, $h$.
The Higgs boson will
certainly be discovered, if it exists, at Fermilab or
 the LHC.  However, discovery is
not enough.  Both the LHC and
a high energy $e^+e^-$ linear collider will be needed
in order to completely elucidate the properties of a Higgs boson and to
verify its role in the origin of mass.

The Standard Model with a single Higgs boson is unsatisfactory, however, 
in several respects, among them the lack of unification of the gauge coupling
constants at high energy
and the emergence of quadratic divergences in the Higgs boson
 mass 
renormalization at 1-loop.  Both of these problems are solved 
by the introduction of supersymmetry at the TeV scale.  The minimal 
supersymmetric model has a large number of new particles, many of which 
can be discovered at the LHC.  Determining that these particles 
correspond to an underlying supersymmetric model, however,
 requires input from both
the LHC and a linear collider.  Furthermore, the scalar partners of leptons
will be extremely challenging to observe at the LHC.

Finally, we consider the role of the top quark.
Precision measurements of the $t {\overline t} h$
Yukawa coupling, $g_{tth}$, and of the top quark mass, $M_t$,
 can help to restrict models
with new physics in the top quark sector.  The top quark mass is
a fundamental parameter for precision electroweak measurements and is
vital for determining the consistency of the Standard Model.  We
summarize the expected precisions for $g_{tth}$ and $M_t$ at the LHC
and an LC.

The theme is clear:  the physics requires a high energy $e^+e^-$
collider in addition to the LHC in
order to complete our understanding of the TeV energy scale.

\section{Higgs Physics}

The Higgs mechanism implies the existence of a scalar particle, the
 Higgs boson, $h$.  The mass of the Higgs
boson is a free parameter in the theory,
 but the couplings to fermions and gauge bosons,
along with  the  Higgs self-couplings, are fixed, allowing for
definitive experimental searches.   
The current direct search limit from LEP is $M_h>114~GeV$, while
precision measurements suggest $M_h<193~GeV$\cite{ichep_ew}.
 A Higgs boson in this mass region can, with sufficient 
luminosity, be discovered at the Fermilab Tevatron\cite{carena}, 
while the LHC will find a Higgs
boson of any mass below $1~TeV$\cite{atlastdr}.   In
order to verify that this particle is the source of mass, we must do more
than simply observe it.  We must:
\begin{itemize}
\item
Measure the Higgs couplings to fermions and gauge bosons.  
In the Standard Model, the couplings to fermions, $g_{f { f}h}$,
 and to gauge bosons,
$g_{WWh}$, are proportional to mass:  
\begin{eqnarray}
g_{f {f}h}&=&{m_f\over v}\\
g_{WWh}&=&gM_W.
\end{eqnarray}
\item
Measure the Higgs self-couplings.  The Higgs self-couplings  are proportional
to the Higgs mass and arise from the potential:
\begin{equation}
V={M_h^2\over 2}h^2+{M_h^2\over 2 v}h^3+{M_h^2\over 8 v^2}h^4~.
\end{equation}
\item
Verify the Higgs spin- parity assignments, $J^{PC}=0^{++}$.
\end{itemize}

\begin{figure}
\begin{center}
\rotatebox{90}{
\includegraphics[width=.6\textwidth]{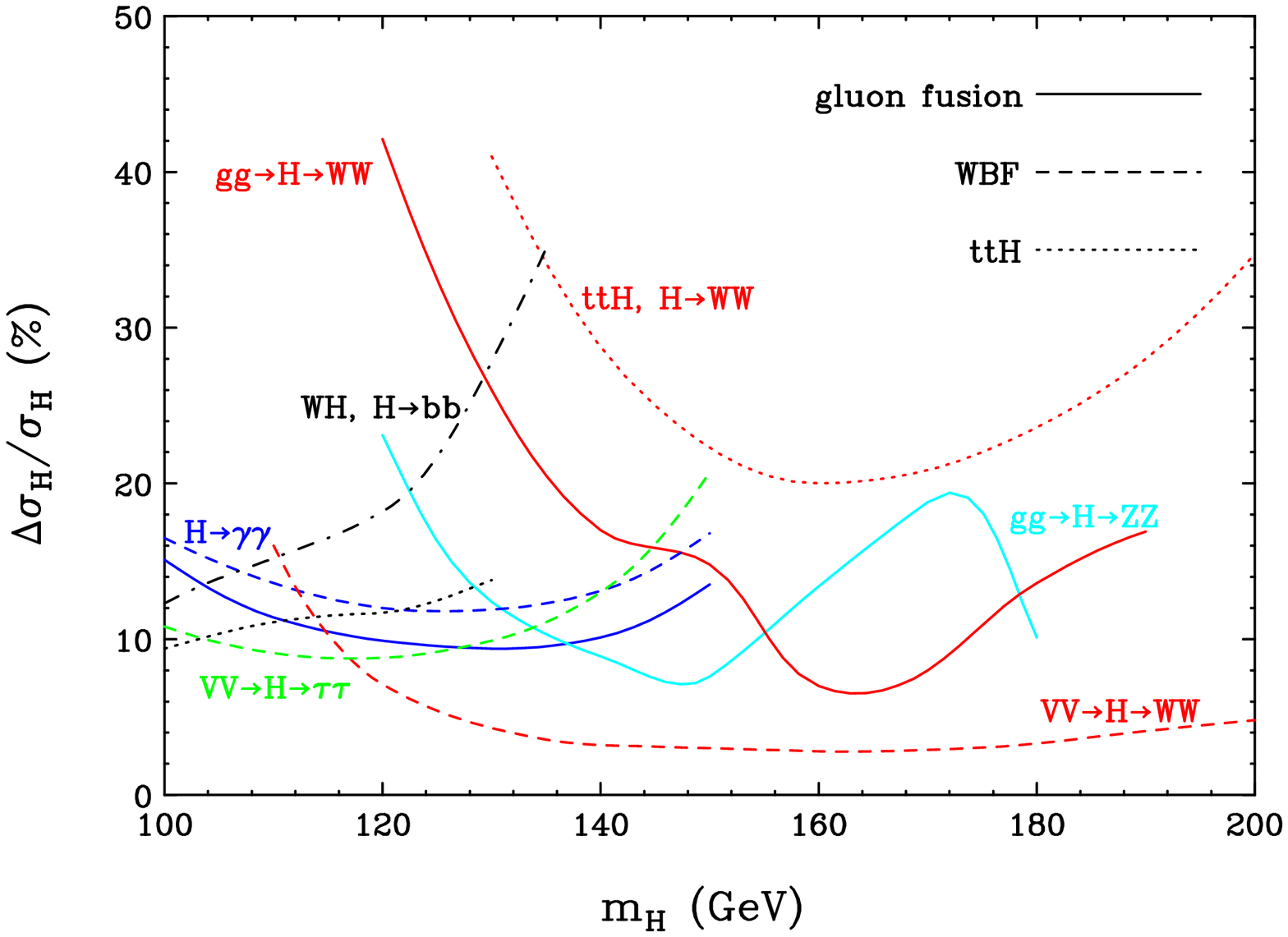}
}
\end{center}
\caption[]{ Relative error on $\sigma \cdot
BR$ at the LHC with ${\cal L}=200~fb^{-1}$.
The $t{\overline t}h, h\rightarrow WW$ and $Wh,h\rightarrow b {\overline b}$
curves assume ${\cal L}=300~fb^{-1}$\cite{zep_brpap}.}
\label{brs_fig}
\end{figure}

The LHC can measure combinations of Higgs coupling constants arising
in the products of production cross sections multiplied by Higgs
branching ratios, $\sigma \cdot
BR$, as shown in Fig. \ref{brs_fig}.  The
experimental accuracy depends sensitively on the Higgs boson
mass.  For $M_h >200~GeV$, only the $h\rightarrow WW$ and
$h\rightarrow ZZ$ channels are observable. 
For a Higgs boson below the $ZZ$ threshold, more decay modes
are accessible.  The measured values of $\sigma \cdot BR$ depend on
parton distribution functions and various combinations of
Higgs couplings  to fermions and gauge bosons.
  By combining the different
channels, ratios of Higgs boson partial decay widths can be extracted
and many of the uncertainties cancel.  
The accuracy
is typically 10-20$\%$\cite{zep_brpap} and only a subset of the Higgs couplings
can be obtained in this manner.

\begin{figure}
\begin{center}
\includegraphics[width=.6\textwidth]{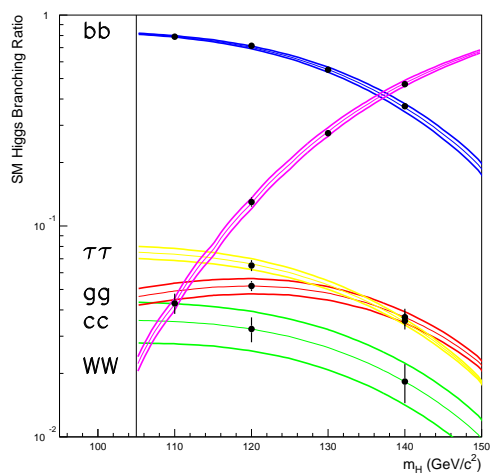}
\end{center}
\caption[]{Higgs branching ratios from $e^+e^-\rightarrow Zh$ 
at $\sqrt{s}=350~GeV$ with ${\cal L}
=500~fb^{-1}$.  The bands represent theory
uncertainty\cite{batt_pap}.}
\label{batt_brs}
\end{figure}

A linear collider benefits from a well determined initial state
with fixed quantum numbers and energy. The linear collider will measure
$e^+e^-\rightarrow Zh$ with a large rate ($\sim 40,000~ h$'s per year).
If the Higgs boson is kinematically accessible, 
an LC can observe the Higgs boson independent from the
Higgs decay pattern using missing mass techniques.
 The Higgs 
branching ratios can then be measured in all channels with a typical
precision of $2-10\%$ as shown in Fig. \ref{batt_brs}\cite{batt_pap,picc}.  
  The data points in Fig. \ref{batt_brs} 
correspond to an integrated luminosity of ${\cal L}=500~fb^{-1}$
 at a center of
mass energy, $\sqrt{s}=350~GeV$.
  The bands correspond to the theory error, of which the
largest component comes from uncertainties on the $b$ quark mass.
 The cross section for $e^+e^-
\rightarrow Zh$ decreases by roughly a factor of two as the center
of mass energy is increased from $\sqrt{s}=350~GeV$ to $\sqrt{s}=500~GeV$
and so the more precise measurements 
of Higgs branching ratios
are obtained at the lower energy.

A precise measurement of 
the mass of the Higgs boson is interesting primarily for comparison
with the value extracted indirectly from electroweak observables.  At
the LHC, the Higgs mass is extracted from reconstruction of the decay
products in the channel, $h\rightarrow \gamma\gamma$.  For $M_h<150~GeV$
and an integrated luminosity  of ${\cal L}=300~fb^{-1}$,
 a precision of $\delta M_h\sim 100~MeV$ can be obtained\cite{conway}.
  A more precise
value can be obtained at the linear collider 
where a precision of $\delta M_h \sim 50~MeV$ can be found for
$M_h=120~GeV$ at $\sqrt{s}=350~GeV$ and ${\cal L}=500~fb^{-1}$\cite{conway}.

\begin{figure}
\begin{center}
\includegraphics[width=.5\textwidth]{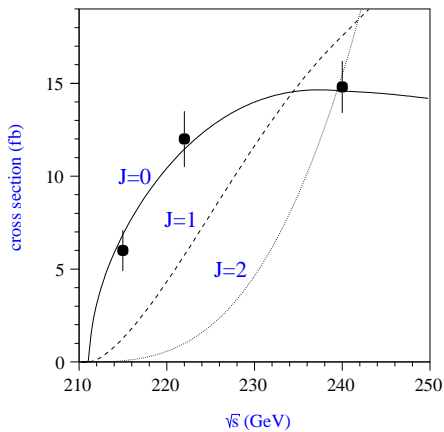}
\end{center}
\caption[]{Dependence of the $e^+e^-\rightarrow Zh$ cross section on
the spin of the Higgs boson\cite{miller}.  The error bars correspond to
$20~fb^{-1}$/point.}
\label{spin_fig}
\end{figure}

The Higgs mechanism requires that the Higgs boson be a spin 0 particle with
positive charge and parity.  The threshold dependence
 of the $e^+e^-\rightarrow Zh$ cross section depends sensitively on the
spin of the Higgs boson and can be measured with a modest amount of
integrated luminosity. The data points shown in Fig. \ref{spin_fig}
correspond to $20~fb^{-1}$ per point \cite{miller} and clearly distinguish
between spin 0,1, and 2 assignments for the Higgs boson.  The angular
dependence of the $Z$ decay products in $e^+
e^-\rightarrow Zh$
are also sensitive to the Higgs CP
assignments.\cite{choi}

\section{Supersymmetry (SUSY)}

If low mass scale ($<1~TeV$)
supersymmetry exists, some indications will be seen in the
first $10~fb^{-1}$ of data at the LHC through a large excess in the
missing $E_T$ plus jets signal.
The task of the LHC and LC will  then be to
untangle the supersymmetric spectroscopy
 and to verify that the new particles are indeed
the result of an underlying supersymmetric theory.  

\subsection{SUSY Higgs Sector}
\begin{figure}[t,b]
\begin{center}
\includegraphics[width=.6\textwidth]{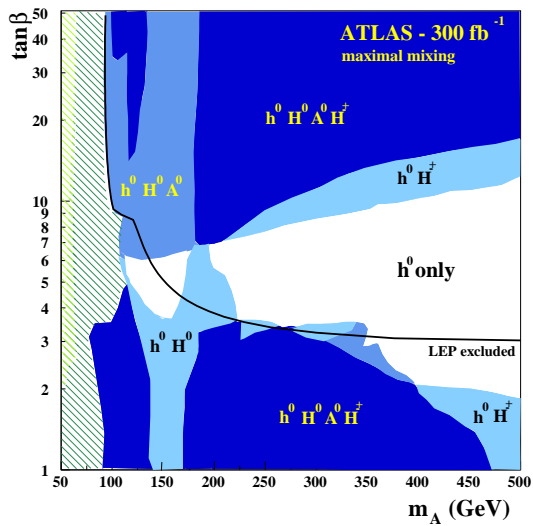}
\end{center}
\caption[]{Supersymmetric Higgs parameter space accessible at the LHC with
${\cal L}=300~fb^{-1}$.  In the white region, only a single Higgs boson can be
observed, with couplings close to those of the SM.\cite{gianotti}}
\label{wedge_fig}
\end{figure}

The Higgs sector of a supersymmetric model is very predictive, with an
upper limit on the neutral Higgs boson mass of around $130~GeV$ and all masses
and couplings predicted at tree 
level
in terms of two parameters\cite{carena}. 
 These parameters are usually taken to be
$M_A$, the mass of the pseudoscalar Higgs boson, and $\tan\beta$,
the ratio of neutral Higgs boson vacuum expectation values.
In order to claim discovery of a supersymmetric Higgs sector, it will
be necessary to observe at least two of the five
Higgs bosons of a SUSY theory with the
predicted properties.  The LHC can discover the lightest neutral Higgs
boson in  all regions of parameter space in
the minimal supersymmetric model,
 but for a significant portion of the parameter space
it cannot observe the other Higgs bosons.  This is the decoupling regime, 
 shown as the white region in Fig. \ref{wedge_fig}\cite{gianotti},
where the lightest
Higgs boson has Standard Model couplings and the other Higgs particles are
heavy.

The ability of an LC to make precision measurements of the Higgs couplings
will be critical in probing the decoupling region of a supersymmetric model.
Deviations in Higgs boson branching ratios from their SM values can provide
information on the heavy Higgs bosons of a SUSY model, even when these
particles
are kinematically inaccessible.  With ${\cal L}=1000~fb^{-1}$, an LC
will be sensitive to the pseudoscalar Higgs mass up to approximately
$600~GeV$\cite{batt_pap}.

\begin{figure}[t,b]
\begin{center}
\includegraphics[width=.6\textwidth]{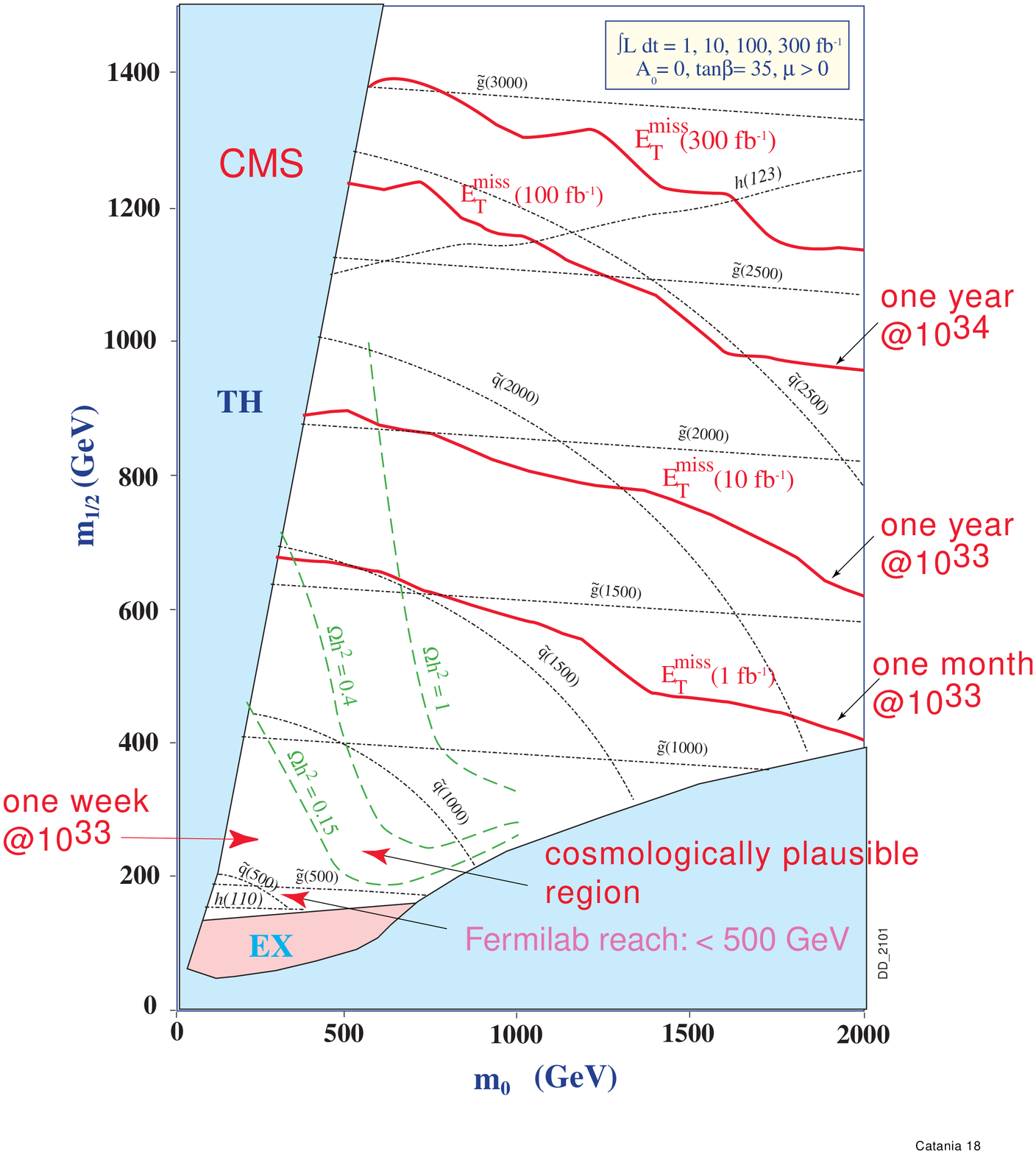}
\end{center}
\caption[]{$5\sigma$ discovery reach for supersymmetric
particles with the CMS detector
at the LHC\cite{cmssusy}.}
\label{cms_fig}
\end{figure}

\subsection{SUSY Partners}
Strongly interacting SUSY particles will be observed with large rates at the
LHC.  The classic signal, jets with missing $E_T$, will arise 
predominantly from squark and gluino production.  The squarks and
gluinos then decay to ordinary quarks and gluons, along with the
lightest SUSY particle (LSP).
The region of SUSY parameter space accessible at the $5\sigma$ significance
level at the LHC is shown in Fig. \ref{cms_fig} 
\cite{cmssusy}.  After a year of
running at the design luminosity, SUSY mass scales in the $1-2~TeV$ region
can be observed.  

It will be difficult to untangle the details
of the SUSY signals, however, since 
the signatures depend sensitively on the details of the specific
supersymmetric model.
The challenge is then to make precision measurements
of masses and couplings.
By measuring cascade decays of SUSY particles,
 some mass differences can be measured at the LHC, 
with  the precision being limited by the unknown mass of the lightest
SUSY particle (LSP)\cite{paige}.  

The linear collider can make precise measurements of particle masses by
scanning threshold cross sections.  If the particles are within the kinematic
reach of the linear collider, they will be pair produced and the masses
measured unambiguously\cite{grannis}.
In order to verify that the new particles observed correspond to
a supersymmetric model, the full spectrum of superpartners must be observed.
For example, both the ${\tilde e}_L$ and the ${\tilde e_R}$ 
scalar partners of the
electron must be seen.  The polarization capabilities of a linear collider
are critical for this measurement.

\section{The Top Quark}

The  origin of the
fermion mass spectrum in the Standard Model is not well understood. 
 A precise
measurement of the top quark mass is of interest for understanding
the source of fermion masses.   Since the coupling of the Higgs
boson to a $t {\overline t}$ pair is proportional to $M_t$,
a  deviation of the top
quark Yukawa coupling, $g_{tth}$,
from its Standard Model value would be a signal for
new physics.
\begin{figure}
\begin{center}
\includegraphics[width=.6\textwidth]{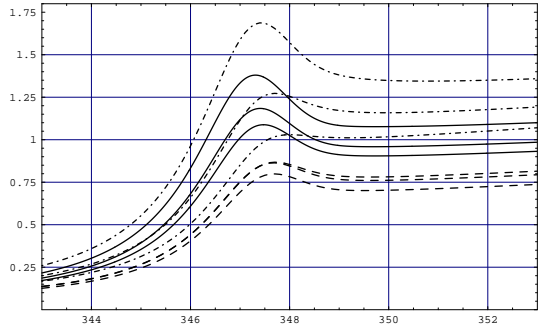}
\end{center}
\caption[]{The threshold cross section for $e^+e^-\rightarrow
t {\overline t}$ as a function of the center of mass energy
in GeV at LO (dashed-dotted), NLO(dashed), and NNLO(solid), for three values
of the arbitrary renormalization scale\cite{oleg}.} 
\label{tthresh}
\end{figure}

The top quark Yukawa coupling
can be measured through $t{\overline t} h$ production,
which has a small rate, but a spectacular signature.  
At the LHC, the signal
and background for $t {\overline t}h$ 
production  have the same shape and there is a $15-20\%$ scale 
uncertainty from the arbitrary renormalization scale.  With
an integrated luminosity of $~{\cal L}=300~fb^{-1}$, the
LHC can obtain an accuracy of\cite{lhctop},
\begin{equation}
{\delta g_{tth}\over g_{tth}}\sim 16\%,~~~~~{\hbox{LHC}}.
\end{equation}

The signature for $t {\overline t}h$
production is cleaner at an $e^+e^-$ linear collider, but 
the rate is limited by phase space and requires 
$\sqrt{s}\sim 700-800~GeV$ in order to be observable.
With a large integrated luminosity of  $ {\cal L}=1000~fb^{-1}$,
a linear collider at $\sqrt{s}=800~GeV$ can obtain\cite{eetth}
\begin{equation}
{\delta g_{tth}\over g_{tth}}\sim 6.5\%,~~~~~{\hbox{LC}}.
\end{equation}
At $\sqrt{s}=500~GeV$, the accuracy is significantly less.

At the LHC with ${\cal L}=50~fb^{-1}$,
 the top quark mass can be measured by reconstruction of the decay
products with a precision\cite{atlastdr} 
\begin{equation}
\delta M_t\sim 1-2~GeV, ~~~~~{\hbox{LHC}}.
\end{equation}
A linear collider can measure the top quark mass quite precisely with a
threshold energy scan with a relatively small amount of luminosity,
${\cal L}=40~fb^{-1}$,
\begin{equation}
\delta M_t \sim 200~MeV,~~~~~{\hbox{LC}}.
\end{equation}
The dependence of the $e^+e^-\rightarrow
t {\overline t}$ threshold cross section is
shown in Fig. \ref{tthresh}.  The cross section has been calculated
to NNLO and with the proper definition of the top quark mass the
perturbation theory is well defined.
The QCD effects are well understood and the scale uncertainty is roughly
$20\%$\cite{oleg}.

The precise value of the top quark mass is of importance for 
the extraction of the Higgs boson
mass from precision electroweak measurements\cite{baur}. 
 Combined with a precise measurement of
$M_W$, a value for the Higgs mass can be extracted indirectly.
Consistency of the SM requires that this prediction
agree with the direct measurement
of $M_h$.  Assuming $M_{h}=115~GeV$ and a measurement of $\delta M_W=15~MeV$,
the LHC indirect measurement of the Higgs boson
mass will have an error, 
$\delta M_h\sim 100~MeV$.  Similarly, assuming $\delta M_W=10~MeV$, the
linear collider will indirectly measure the Higgs mass with an accuracy of 
$\delta M_h\sim 50~MeV$.

The precision measurement of the top quark is particularly interesting
in the case of a supersymmetric model.  Fig. 7 
demonstrates the allowed parameter space from measurements of
$M_W$ and $M_t$ at the LHC and shows the improvement at a LC running
at high luminosity at the $Z$ pole (giga-Z).  In this case, the
precision measurements will clearly discriminate between the SM
and the MSSM\cite{hein}.
\begin{figure}
\begin{center}
\includegraphics[width=.6\textwidth]{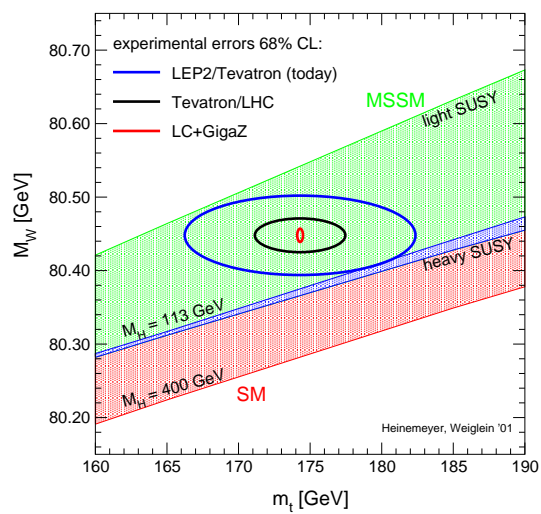}
\caption[]{Indirect measurement of $M_h$ in the SM and in the
minimal supersymmetric model\cite{hein}.}
\end{center}
\label{prec_fig_sh}
\end{figure}

\section{Conclusion}

The physics program of a high energy $e^+e^-$ linear collider, in conjunction
with that of the LHC, can vastly expand our understanding of the TeV
scale.  Preliminary studies of possible run plans at a linear
collider \cite{grannis} 
indicate that with $1000~fb^{-1}$, most of 
the physics goals described in this note
should be obtainable.

\end{document}